\documentstyle[11pt,pasp,twoside,epsf]{article}
\def\edcomment#1{\iffalse\marginpar{\raggedright\sl#1\/}\else\relax\fi}
\reversemarginpar

\begin{document}
\title{Accretion Flows in X--Ray Binaries and Active Galactic Nuclei}
 \author{Chris Done}
\affil{Department of Physics, University of Durham, South Road,
Durham, DH1 3LE, UK}

\begin{abstract} Hard X--ray emission is ubiquitous in accreting black 
holes, both in
Galactic binary systems and in Active Galactic Nuclei. I review the
different spectra which can be seen from these systems, and possible
ways of producing this emission. X--ray reflection should give an
observational test of these scenarios, but we need
better models before this can give an unambiguous diagnostic
of the accretion geometry.

\end{abstract}

\section{Introduction}

Accretion onto a black hole is observed to produce copious X--ray
emission. For the Galactic binary systems, then at high mass accretion
rates (luminosity close to the Eddington limit, ${\rm L_{Edd}}$) the
spectra are dominated by a soft component at $kT\sim 1$ keV which is
strongly (very high state) or weakly (high state) Comptonized by low
temperature thermal (or quasi--thermal) electrons with $kT\sim
5$--$20$ keV (e.g. Gierli\'nski et al. 1999; {\.Z}ycki, Done \& Smith
2001; Kubota et al. 2001). There is also a rather steep power law tail
($\Gamma\sim 2$--$3$) which extends out beyond $511$ keV in the few
objects with good high energy data (e.g. Grove et al. 1998). At lower
mass accretion rates, below $\sim 0.02$--$0.03 {\rm L_{Edd}}$
({\.Z}ycki, Done \& Smith 1998; Gierli\'nski et al. 1999), there is a
rather abrupt transition when the soft component drops in temperature
and luminosity. These low state spectra are dominated by the hard
component with $\Gamma < 1.9$, rolling over at energies of $\sim 150$
keV (see e.g. the reviews by Tanaka \& Lewin 1995; Nowak 1995).  This
spectral form continues to the very low luminosities of the quiescent
state ($\sim 10^{-4} {\rm L_{Edd}}$ e.g. Kong et al. 2000).

In AGN, nearby Seyfert galaxies generally have hard X--ray spectra
which are rather similar to the low state spectra from Galactic black
holes (Zdziarski et al. 1995), 
while the (probably higher mass accretion rate) Narrow Line
Seyfert 1 galaxies 
and radio--quiet quasars look more similar to the
high/very high state spectra (Pounds, Done \& Osborne 1995;
Sulentic et al. 2000; Kuraszkiewicz et al. 2000), 
as do the ultraluminous off--nuclear
sources in nearby galaxies (compare spectra in Makashima et
al. 2000 with those of e.g {\.Z}ycki et al. 2001).

\section{Hard X--ray Emission Mechanisms}

The low state (and quiescent) continuum spectra 
are adequately described by thermal Compton
scattering of UV/soft X--ray seed photons in an X--ray hot plasma
(e.g. Sunyaev \& Treumper 1979)
which has temperature
$\sim$100 keV, and optical depth $\sim$1 
(e.g. Gierli\'nski et al. 1997).  
This is often modelled by a power law with exponential cutoff, but
true Compton spectra can be subtly curved from any anisotropy of
the seed photons (e.g. if they come from a disk: Haardt \& Maraschi
1993) rather than a power law. The high energy cutoff 
is also much sharper than 
exponential at these temperatures
(Poutanen \& Svensson 1996).

By contrast, the high/very high state continuum spectra cannot be
fit by seed photons from a disk Comptonized by a single temperature
plasma. The data require complex curvature, best described by a
two component electron distribution, which
is quasi--thermal at low energies, and a power law at higher energies
(Poutanen \& Coppi 1998; Gierli\'nski et al. 1999; Wilson \& Done 2001).

There are several proposed mechanisms to heat the electrons.  The
advective flow models assume that the accretion energy is given mainly
to the protons and is mostly carried along with the flow with only a
fraction being given to the electrons via Coulomb collisions (e.g. the
review by Narayan et al. 1998). This gives a natural spectral switch
at a few percent of Eddington as at high densities (i.e. high mass
accretion rates) the electrons efficiently drain energy from the
protons and the flow collapses back into an Shakura--Sunyaev disk
(e.g. Esin, McClintock \& Narayan 1997).  These models can easily
give continuity of spectral form between quiescence and low state, and
give the spectral switch to the high state, but {\it cannot} produce
the X--ray emission seen in the high states.

An alternative mechanism for the low state emission is magnetic
reconnection, releasing the accretion energy in a flaring corona above
the disk.  This seems plausible as the disk viscosity is known to be
connected to an MHD dynamo (Balbus \& Hawley 1991). However, 
explaining the similarity of the low state and 
quiescent spectra may then be difficult as it is
likely that the magnetic viscosity cannot operate at extremely
low accretion rates (Gammie \& Menou 1998).  Another problem is that
the only known potential mechanism for the emission in the 
high states is magnetic
reconnection. The spectra change dramatically between the
high and low states, which seems difficult to explain
if they are both powered by
the same mechanism. The radiation pressure instability probably 
does not offer
a potential spectral switch as it is suppressed by a magnetic corona
(Zdziarski \& Svensson 1994). 

The increased Comptonization in the very high state could be caused
either by the inner disk becoming effectively optically thin (Shakura
\& Sunyaev 1973), or by incomplete thermalization in the magnetic
reconnection process leading to a hybrid (thermal plus nonthermal
tail) electron distribution (Poutanen \& Coppi 1998).  
These possibilities are shown in Figure 1.

\section{Observational Tests using X--ray Reflection}

The potential emission mechanisms shown in Figure 1 imply very
different geometries. X--ray reflection is an observable diagnostic of
the geometry, as the amount of reflection depends on the solid angle
of optically thick material (disk) as seen from the hard X--ray
source, while the amount of relativistic smearing of the reflected
spectral features 
shows how far the material extends down into the gravitational
potential (see e.g. the review by Fabian et al. 2000). 

The solid angle subtended by the reflecting disk in AGN seems to be
correlated with spectral index (Zdziarski et al. 1999), and also with
the amount of relativistic smearing (Lubi\'nski \& 
Zdziarski 2001).  However,
these observational results are currently controversial. The broad
line profiles are sensitive to details of the ASCA (re)calibration
(Weaver, Gellbord \& Yaqoob 2001), and to the extent of any narrow
line component from material further out such as the molecular torus
or broad line region. The correlation of reflection with the
illuminating power law index could be affected by details of the 
spectral decomposition (and background subtraction) for faint AGN
(Nandra et al 2001). But this is 
{\it not} a problem with the Galactic black holes, 
as their typical fluxes are much higher. Yet
these again show the {\it same} correlation between the spectral index
and amount of reflection (Ueda et al. 1994; {\.Z}ycki et al. 1999;
Zdziarski et al. 1999; Gilfanov et al. 1999). They also show a
correlation of the amount of smearing of the spectral features with
amount of reflection/spectral index (Gilfanov et al 2000), and again
this {\it cannot} be due to any of the problems seen in AGN (the data
are from RXTE rather than ASCA, and there is no molecular
torus to produce a narrow line component).  These correlations in the
Galactic binaries {\it must} be
telling us something real about the X--ray accretion geometry.

\section{Reflection Constraints on the Accretion Geometry ?}

These results can rather naturally be
explained in an advective flow/truncated disk geometry if the inner
disk radius moves inwards as a function of mass accretion rate. This
automatically gives more relativistic broadening. If this is
accompanied by an increasing overlap between the disk and advective
flow then this increases the solid angle subtended by the disk, so
increasing the amount of reflection. It also increases the amount of
seed photons which are intercepted by the X--ray hot plasma, so leads
to a steeper Comptonized spectrum (Poutanen et al. 1997; Zdziarski et
al. 1999; {\.Z}ycki et al. 1999; Gilfanov et al. 1999). More evidence
for this interpretation comes from the fact that a hole in the
quiescent disk may be required to make the long inter--outburst
timescale in the black hole transient systems (Dubus et al. 2001), 
while the {\it
variable} frequency QPO seen in these sources argues for a variable
radius for the inner disk edge.

However, this is not a unique interpretation of these
correlations. The magnetic flares model can give similar correlations
if the flares are accompanied by relativistic outflow away from the
disk: the faster the outflow the more the emission is beamed away from
the disk so the lower the reflection and the lower the seed photon
flux seen by the X--ray plasma (Beloborodov 1999). 

Alternatively, the magnetic flares can be static if they irradiate the
disk to an extent that the X--ray heating substantially changes the
vertical structure of the disk. Material at the top of the disk is
heated by the illumination, and so can expand, lowering its density
and increasing its ionization.  This self--consistent density
structure is especially important for the 
low state systems, as the intense hard spectral illumination
causes an ionization instability in the disk. This results in an abrupt
transition from a hot, intensely ionised skin (where X--ray heating is
balanced by Compton and bremsstrahlung cooling) to low temperature,
high density, low ionization material (where X--ray heating is
balanced mainly by line cooling: see e.g. Nayakshin, Kazanas \&
Kallman 2000).

Models with this X--ray illuminated disk structure can show an
apparent correlation between the spectral index and amount of
reflection as inferred from using the standard constant disk
ionization state reflection models
(Done \& Nayakshin 2001a), although the models
depend on (unknown) physics of magnetic flares, e.g. any {\it
sideways} expansion of the X--ray heated patch of disk underneath a
reconnecting region (Done \& Nayakshin 2001b).

\begin{figure}
\plotone{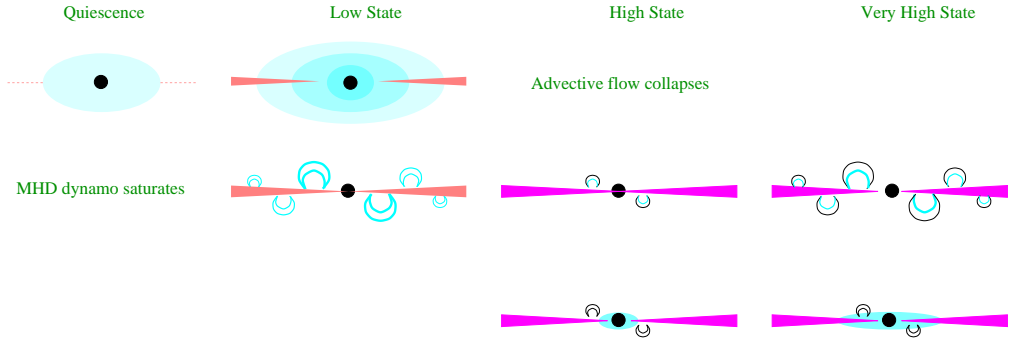}
\caption{Potential X--ray emission mechanisms.
In quiescence the disk is NOT in steady state (dotted line). 
The MHD dynamo probably cannot operate, so advective
flows may be the only feasible way to power the X--ray emission.  In
the low state the accretion flow is probably in (quasi)steady state
and the MHD dynamo works. If most of the reconnection takes place
above the optically thick material then this can power the observed
X--ray emission.  Alternatively, if quiescence if powered by an
advective flow then maybe the low state is also.  In the high and very
high states the only serious contender for the hard power law tail is
magnetic reconnection leading to a non--thermal electron distribution
(indicated by the black loops), while the thermal electrons could be
part of the same mechanism (grey loops) or could be associated with
the inner disk.}
\end{figure}

\section{Reflection in the high/very high states}

The complex curvature of the high energy spectrum (sum of thermal and
non--thermal Compton components) seen in the high/very high states
makes it difficult to determine the reflected continuum. The Galactic
black hole systems are further complicated by the dominance of the
disk spectrum in the observed X--ray bandpass. 
Relativistic smearing (e.g. Ebisawa et al. 1991) and weak
Compton scattering in the upper layers of the disk (e.g. Ross, Fabian
\& Mineshige 1992) should broaden the
optically thick disk spectrum. However, much stronger Compton scattering
is implied by data,
which probably dominates over any relativistic broadening in the thermal
disk spectrum ({\.Z}ycki, Done \& Smith 2001; Kubota et al. 2001).
This (poorly known) disk spectrum is the seed
photon source, so there is a low energy cutoff in the Compton
continuum spectrum. 

Despite these uncertainties, there are some robust results. Reflection
is {\it always} present and is {\it always} highly ionised ({\.Z}ycki et
al. 1998; 2001; Gierli\'nski et al. 1999; Wilson \& Done 2001). 
However, the amount of reflection is not easy to constrain:
strong ionization means 
that there can be strong Comptonization of the
reflected features which makes reflection much harder to identify
(e.g. Ross, Fabian \& Young 1999). 
Radial ionization gradients can also be important:
there could be material in the inner disk which has even higher
ionization than the material which we see. This would be
completely reflective so would lead to an underestimate of the solid
angle of the disk ({\.Z}ycki et al. 1998). 

\section{Concluding Questions}

\noindent $\bullet$ {\bf Theory Issues}: Can the Balbus--Hawley
viscosity work in quiescent disks ? Can it transport most of the
accretion energy into an optically thin corona to power the low state
emission ?  What happens to the vertical structure of the disk under
such intense illumination ? For the alternative accretion geometry,
can the advective flow models exist i.e. can the accretion energy
really be given mainly to the protons and are (slow) Coulomb
collisions the only process by which the electrons can gain this
energy ?  And what controls the tranistion
radius between the disk and advective flow ?  How much Comptonisation
can be produced in the disk at high mass accretion rates ?  And a
topic which I completely ignored: what role does the jet play in the
X--ray emission from any of the spectral states~?

\noindent $\bullet$ {\bf Modelling Issues}: We need to fit proper
reflection models to the data (not smeared edges and broad gaussian
lines!). These models should include the self--consistently calculated
vertical ionization structure of the X--ray illuminated accretion
disk, especially in low state spectra where the thermal ionization
instability is expected to develop. They should also include Compton
upscattering of the reflected emission in the hot, ionised upper
disk. And we need to use proper Comptonisation models of the continuum
(including both high and low energy cutoffs) rather than a power law
with exponential rollover!

\noindent $\bullet$ {\bf Data issues} We need high signal to noise
spectra of all states with good energy resolution around the iron K
features, and broad bandpass to disentangle the continuum from the
reflected emission. This is especially important for the rather faint
spectra from AGN and quiescent Galactic black holes.

Chandra, XMM and {\it especially} ASTRO-E2 will address the
observational deficiencies. The challenge is now for both theorists and
observers to develop and fit physical models to these X--ray spectra.

\end{document}